# Elastic energy of multi-component solid solutions and strain origins of phase stability in high-entropy alloys

Reza Darvishi Kamachali[*], Lei Wang

*Federal Institute for Materials Research and Testing (BAM), Unter den Eichen 87, 12205 Berlin, Germany*

The elastic energy of mixing for multi-component solid solutions is derived by generalizing Eshelby's sphere-in-hole model for binary alloys. By surveying the dependence of the elastic energy on chemical composition and lattice misfit, we propose a lattice strain coefficient $\lambda^*$. Applying to several high-entropy alloys and superalloys, we found that most solid solution alloys are stable when $\lambda^* < 0.16$, analogous to the Hume-Rothery atomic-size rule for binary alloys. We also reveal that the polydispersity index $\delta$, frequently used for describing strain in multi-component alloys, is directly related to the elastic energy ($e$) with $e = q\delta^2$, $q$ being an elastic constant. Furthermore, the effects of (i) the number and (ii) the atomic-size distribution of constituting elements on the phase stability of high-entropy alloys were quantified. The present derivations open for richer considerations of elastic effects in high-entropy alloys, offering immediate support for quantitative assessments of their thermodynamic properties and studying related strengthening mechanisms.

Phase stability is a fundamental topic in the theory of alloy formation. The seminal works by William Hume-Rothery and colleagues [1,2] summarized four empirical rules for formation of solid solutions, known as Hume-Rothery rules. The first rule is on the effect of lattice distortion, stating that the solubility in a binary solid solution is limited when the atomic-size mismatch is above 15%. The importance of lattice distortion in interpreting the phase stability in multi-principal element alloys, also termed as compositionally complex alloys or high-entropy alloys (HEAs), has been discussed in many recent studies (see for instance [3–8] and references therein). The lattice distortion is computed with reference to the solution lattice constant $\bar{a}$, usually described by Vegard's law [9]

$$\bar{a} = \sum_{i}^{N} X_i a_i = X_1 a_1 + X_2 a_2 + \cdots + X_N a_N, \quad (1)$$

based on which the lattice strain coefficients can be written as

$$\lambda_{ih} = \frac{1}{\bar{a}} \left( \frac{\partial \bar{a}}{\partial X_i} \right)_{\substack{X_j \\ i \neq j \neq h}} = \frac{a_i - a_h}{\bar{a}} \approx \frac{r_i - r_h}{\bar{r}} \quad (2)$$

where $\sum_{i}^{N} X_i = 1$ and $\bar{a} \approx \bar{r} = \sum_{i}^{N} X_i r_i$. Here $X_i, a_i, r_i$ are the solute fraction, the lattice constant and the atomic radius of $i$ species, respectively, and $h$ indexes the solvent (host) species. Here the choice of the host element $h$ is arbitrary and does not influence the elastic considerations, as will be shown in the following.

In studying the extent of lattice distortion in multi-component alloys, the parameter

$$\delta = \sqrt{\sum_{i}^{N} X_i \left( \frac{r_i - \bar{r}}{\bar{r}} \right)^2} \quad (3)$$

has been popularly in use. The $\delta$ parameter (also known as polydispersity index) is adopted from the discussions on crystallization where in the limit of $\delta < 0.06$, a stable crystal is suggested to form from a liquid phase [10–12]. For HEAs, the $\delta$ parameter has been successful in distinguishing the solid solutions from amorphous alloys [13–16] but limited in discerning the solid solutions from intermetallic phase forming alloys. It is found that some intermetallic phase forming alloys show small $\delta$ values [16,17].

Despite similarities in the premises, the parameters $\lambda_{ih}$ and $\delta$ are fundamentally different. While $\delta$ gives the standard deviation from the mean atomic radius $\bar{r}$, the lattice strain coefficients $\lambda_{ih}$s are based on the atomic-size mismatches between *pairs* of species with reference to a chosen solvent $h$. As a matter of fact, the Hume-Rothery atomic-size rule is based on the lattice strain coefficient, rather than the $\delta$ parameter, with the criterion $\lambda_{BA} \approx \frac{r_B - r_A}{r_A} > 0.15$ [1,2]. The elastic energy of a solid solution is also computed based on the lattice strain coefficients; Assuming linear elasticity, Eshelby [18] derived the elastic energy of mixing for an isotropic binary solid solution as $e = q\lambda_{BA}^2 X_B(1 - X_B)$ with $q$ being an elastic constant.

[*]Corresponding author: reza.kamachali@bam.de

In spite of their insightful content, the Hume-Rothery atomic-size rule and the Eshelby equation for the mixing elastic energy are limited to binary alloys and cannot be directly applied for studying multi-component solid solutions. To address this limitation, here we have generalized Eshelby's sphere-in-hole approach and derived the elastic energy of mixing for substitutional multi-component alloys (Appendix A). For an elastically isotropic $N$-component random solid solution we obtain:

$$e = q \sum_{\substack{i \neq h}}^{N} \lambda_{ih}^2 (X_i - X_i^2) - 2q \sum_{\substack{i,j \neq h \\ i \neq j}}^{N} \lambda_{ih} \lambda_{jh} X_i X_j . \quad (4)$$

with $q = 2\mu \frac{1+\nu}{1-\nu}$ where $\mu$ and $\nu$ the shear modulus and Poisson ratio of the isotropic solution, respectively. The first sum in Eq. (4) runs over all species except the chosen host species $h$ and the second sum runs over all dissimilar $i$ and $j$ solute species, revealing cross-interactions elastic energies. Unlike the first sum, which is always positive, the second sum can be positive or negative, depending on the sign of the $\lambda_{ih}\lambda_{jh}$. It can be shown that Eq. (4) is independent of the choice of host species $h$, as $\lambda_{ih}$s are antisymmetric $\lambda_{ih} = -\lambda_{hi}$ and $\lambda_{ih} = \lambda_{ik} + \lambda_{kh}$. Similar to Eshelby's work, the current derivations are in the limit of linear elasticity and assuming a random solid solution. Appendix A gives detailed derivations of Eq. (4).

The mixing elastic energy is always positive ($e \geq 0$) and promotes demixing (instability) of a solid solution. This means that $e$ is in competition with the configurational entropic energy ($-T\Delta S^{\text{conf}}$) and the chemical mixing enthalpy ($\Delta H^{\text{chem}}$, when it is negative) [19,20]. Hence, it is clear that the elastic energy plays a critical role in the phase instability of HEAs. To investigate the effect of elastic energy on the stability of a solid solution, we need to study the variation of the elastic energy upon redistribution of solutes as in a chemical decomposition as follows.

For an $N$-component alloy, a large number of decomposition scenarios are possible. Considering the simplest decomposition scenario, i.e., a binary $i$-$j$ substitution with $\Delta X_i = -\Delta X_j$, with a small composition variation $\Delta X_i$, we obtain:

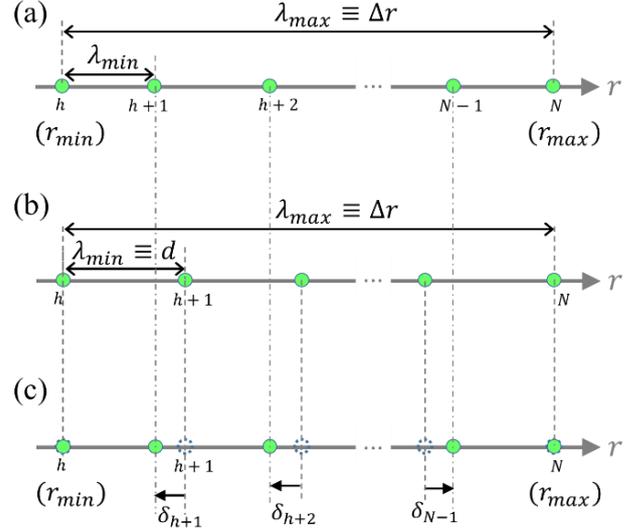

**Figure 1**: The relative atomic-sizes of constituting species of a multi-component alloy. (a) Species of a real alloy arranged based on their atomic-sizes, (b) an idealized case of regularly spaced atomic-sizes for fixed $\Delta r$, and (c) the difference between the real (a) and ideal (b) cases. (b) and (c) are used in deriving Eq. (8). Note that indexes $h+1$, $h+2$, $N-1$ etc. are used to express the next/previous species in the above series arranged by atomic-size. This convention is also used in following to simplify our equations.

$$\left.\frac{\Delta e}{\Delta X_i}\right|_j = q(\lambda_{ih}^2 - \lambda_{jh}^2)$$
$$-2q(\lambda_{ih}^2 - \lambda_{ih}\lambda_{jh})X_i$$
$$+2q(\lambda_{jh}^2 - \lambda_{ih}\lambda_{jh})X_j \quad (5)$$
$$-2q(\lambda_{ih} - \lambda_{jh})\sum_{k \neq h,i,j}^{N} \lambda_{kh} X_k$$

that describes the elastic driving force for a chemical decomposition through the binary substitution of solute atoms —the subscript $j$ in the left side of the equation indicates this $i$-$j$ substitution. See detailed derivation in Appendix B.

Equation (5) gives some key interesting results; while we get $\left.\frac{\Delta e}{\Delta X_i}\right|_j = -\left.\frac{\Delta e}{\Delta X_j}\right|_i$ and the elastic driving force vanishes for $\lambda_{ih} = \lambda_{jh}$, the leading term in this relation turns out to be independent of the alloy composition. All other terms in this equation are weighted by $X_{i/j/k} < 1$. Neglecting those terms, the tendency of the alloy for a chemical decomposition is proportional to $\left.\frac{\Delta e}{\Delta X_i}\right|_j \approx q(\lambda_{ih}^2 - \lambda_{jh}^2)$. Thus, we can identify a lattice strain coefficient:

$$\lambda^* = \sqrt{\lambda_{max}^2 - \lambda_{min}^2} \qquad (6)$$

for a larger value of which a higher tendency for chemical decomposition is expected. Here note that even if considering the composition-dependent terms in Eq. (5), the coefficients $(\lambda_{ih}^2 - \lambda_{ih}\lambda_{jh})$ and $(\lambda_{ih} - \lambda_{jh})$ are proportional to $\lambda^*$ (See Appendix B) and thus $\left.\frac{\Delta e}{\Delta X_i}\right|_j \propto \lambda^*$ still remains a valid conclusion. If we arrange the constituting species of an alloy based on their atomic-sizes, Fig. 1(a), and choose $h$ to be the smallest one $(r_h = r_{min})$, we get $\lambda_{max} = \left|\frac{\Delta r}{\bar{r}}\right|$, with $\Delta r = r_{max} - r_{min}$, and $\lambda_{min} = \left|\frac{r_{h+1} - r_h}{\bar{r}}\right|$. Note that we can also choose $h$ to be the largest atom in the series based on which $\lambda_{min} = \left|\frac{r_N - r_{N-1}}{\bar{r}}\right|$, if it is smaller than $\left|\frac{r_{h+1} - r_h}{\bar{r}}\right|$. In the following calculations we adopt $\lambda_{min} = \left|\frac{r_{h+1} - r_h}{\bar{r}}\right|$ unless otherwise mentioned.

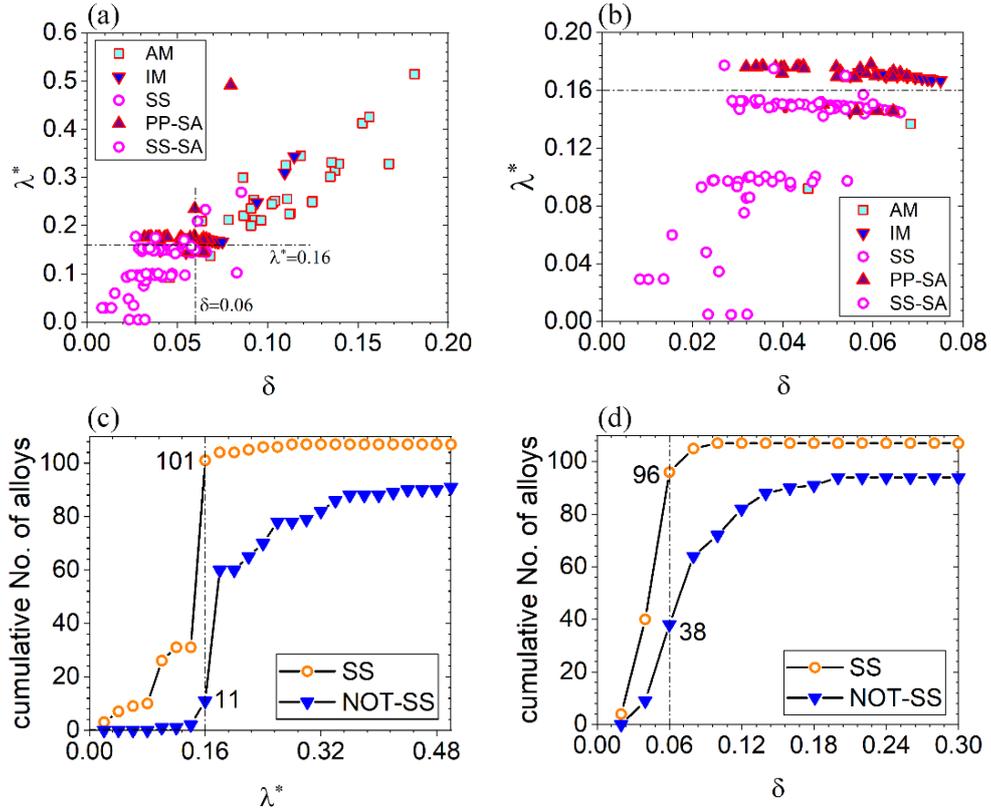

**Figure 2**: $\lambda^*$ representation of several multi-component alloys. (a) $\lambda^*$ and $\delta$ values compared for 201 alloys including 64 solid solution (SS), 21 intermetallic forming (IM) and 33 amorphous (AM) HEAs as well as 43 solid solution superalloys (SS-SA) and 40 precipitate forming superalloys (PP-SA). (b) is a zoom-in of (a). A threshold $\lambda^* < 0.16$ enables a distinct separation between SS and NOT-SS alloys. The cumulative number of alloys for given (c) $\lambda^*$ and (d) $\delta$ show their prediction performance. The sharp jump in (c) indicates the strong predictability of $\lambda^*$ parameter.

Figures 2(a) and (b) show the $\lambda^*$ and $\delta$ values calculated for 201 multi-component alloys. These include 64 solid solution (SS), 21 intermetallic (IM) and 33 amorphous (AM) phase forming HEAs. For comparison, we also applied $\lambda^*$ to 43 solid solution superalloys (SS-SA) and 40 precipitate forming superalloys (PP-SA). The data are collected from [17,21]. Figures 2(a) and (b) reveal that the $\lambda^*$ index sharply separates between the SS (i.e. SS and SS-SA) and other NOT-SS (i.e. IM, AM and PP-SA) alloys. Most notably, the overlap between the SS and NOT-SS alloys, observable on the $\delta$ axis, is prevented on the $\lambda^*$ axis. In the following, we first discuss the significance of the $\lambda^*$ index in identifying SS alloys and then analyze its performance versus the $\delta$ parameter in separating SS and NOT-SS alloys.

Figure 2(c) shows the cumulative number of identified alloys as a function of $\lambda^*$ values. A distinct threshold of $\lambda^* \approx 0.16$ is found below which 101 SS alloys (out of 107 SS alloys studied here) are correctly predicted, i.e. ~94% accuracy. To reconfirm our findings, we applied the $\lambda^*$ index to another set of data on solid solution forming HEA systems, recently reviewed by Steurer [22]. The results are shown in Fig. 3. Using the $\lambda^* < 0.16$ criterion, 65 single-phase HEAs (out of total 82), i.e. ~80%, are predicted. Note that here to calculate $\bar{r}$s we assumed equiatomic alloy compositions with $\bar{r} = \frac{1}{N}\sum_i^N r_i$.

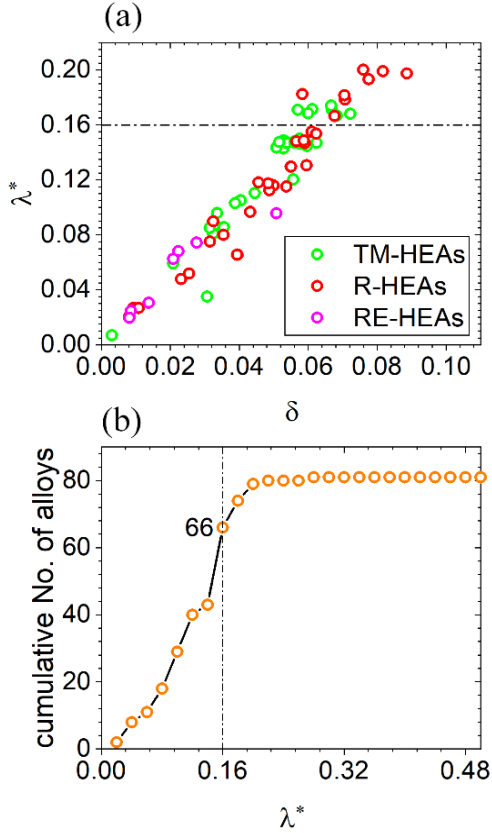

**Figure 3**: $\lambda^*$ of single-phase forming HEA systems. (a) The $\lambda^*$ vs. $\delta$ values (b) The cumulative number of predicted alloys as a function of $\lambda^*$. Here we consider 82 alloy systems as reviewed by Steurer [22] including 39 transition metals (TM), 35 refractory (R) and 8 rare-earth (RE) HEA systems. To calculate the $\lambda^*$ and $\delta$ values we assumed equiatomic alloys.

Eshelby showed that there is a direct relationship between the Hume-Rothery atomic-size rule and the elastic energy for a stable binary solid solution [18]. The $\lambda^* < 0.16$ (16%) criterion, that is very close to Hume-Rothery 15% limit, can be considered as a generalization of the Hume-Rothery rule to multi-component alloys. The physical origin and remarkable performance of the $\lambda^*$ index suggest that it can be used as an indicator for studying the solubility in multi-component alloys.

For most alloys, we found that the $\lambda_{max}$ ($\propto \Delta r$) value plays a dominant role in determining $\lambda^*$, Eq. (6).

This means that the largest atomic-size mismatch, rather than the mean atomic-size mismatch represented by the $\delta$ parameter, can better describe the phase stability. These results also confirm the significance of the geometrical index $\gamma$ previously proposed based on the smallest and largest atoms in a solid solution [21,23,24].

For some alloys, the $\lambda_{min}$ can be large, playing an important role in stabilizing a single-phase solid solution with a large lattice distortion. For instance, for the NbTaTiV HEA [25] we have $\frac{\lambda_{min}}{\lambda_{max}} = \frac{r_{Ti}-r_{Ta}}{r_{Ti}-r_V} \approx 0.23$. As the stability of a phase is always discussed *relatively*, compared to the alternative configurations, it is possible that a phase with a high strain energy can be stabilized when there is no other configuration to relax to. This effect is coined inside the $\lambda_{min}$ value: A large value of $\lambda_{min}$ means that even if $\lambda_{max}$ is large, the most probable chemical decomposition might not result in a significant decrease in the elastic energy and therefore is not elastically promoted.

A major benefit of using the criterion $\lambda^* < 0.16$ is that only 11 NOT-SS alloys (out of 94 NOT-SS alloys studied here) overlap with the 101 correctly predicted SS alloys, i.e., the overlap between SS and NOT-SS alloys becomes very small. These results are shown in Fig. 2(c). The performance of $\delta$ parameter in distinguishing SS from NOT-SS alloys is compared in Fig. 2(d): While the $\delta < 0.06$ predicts the SS alloys fairly well (predicting 96 SS alloys out of total 107 SS alloys studied here), it is clear that a much larger overlap between the SS and NOT-SS alloys is obtained. Figure 2(d) shows that 38 out of 94 NOT-SS alloys overlap with the SS alloys. The results show that other values instead of 0.06 can only worsen the ability of the $\delta$ parameter in distinguishing SS and NOT-SS alloys.

To explain the origin of this limitation of the $\delta$ parameter, one needs to understand why some NOT-SS forming alloys show low $\delta$ values? Interestingly, by further inspection of Eq. (3) in the context of the strain coefficients $\lambda_{ih}$s and elastic energy, Eqs. (2) and (4), we found:

$$e = q\delta^2 \qquad (7)$$

i.e., the $\delta$ parameter in fact uniquely corresponds to the mixing elastic energy of an isotropic multi-component alloy. See the derivations in Appendix C. Equation (7) is a key result of our study, shedding light on the physical meaning of the $\delta$ parameter. For $\delta = 0.06$ and $q \sim 150 - 400$ GPa we immediately see that $e \sim 40 - 100$ meV per atom, that can be a large fraction of the total enthalpy of mixing for some HEAs [26].

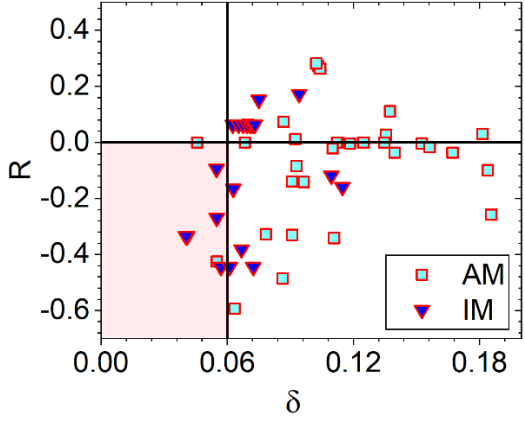

**Figure 4**: The relationship between the deviatoric energy term (second term in Eq. (8)) and the $\delta$ values for IM and AM HEAs. $R = \frac{\text{deviatoric term}}{\text{first term}} = \frac{12\xi(N-2)}{N\Delta r(N+1)}$. The results show that for AM and IM HEAs with R< 0, $\delta < 0.06$.

For a more interpretive understanding of the $e$ and $\delta$ values, we rederived the elastic energy of mixing (Eq. (4)) on the basis of having an *equiatomic* HEA. For an $N$-component equiatomic HEA, $X_i s = \frac{1}{N}$, we obtain (See Appendix D):

$$e = \frac{q\,\Delta r^2}{12\left(r_{min} + \frac{\Delta r}{2}\right)^2}\left(\frac{N+1}{N-1}\right) \\ + \frac{q\,\Delta r\,\xi}{N\left(r_{min} + \frac{\Delta r}{2}\right)^2}\left(\frac{N-2}{N-1}\right) + O(\delta_i^2). \quad (8)$$

Here we conducted a two-step derivation; First, we considered an ideal case in which atomic-sizes are regularly spaced (equidistance), as illustrated in Fig. 1(b), with $d = \frac{\Delta r}{N-1}$, $\lambda_{max} = \frac{(N-1)d}{\bar{r}}$, $\lambda_{min} = \frac{d}{\bar{r}}$ and $\bar{r} = r_{min} + \frac{\Delta r}{2}$. This gives the first term in Eq. (8). Then we computed the additional (deviatoric) elastic energy due to a deviation from this ideal configuration: This gives the second term in Eq. (8). To do so, we introduced deviatoric atomic-size mismatches $\delta_{h+k} = r_k - kd$, as defined in Fig. 1(c). The deviatoric elastic energy scales with a new parameter

$$\xi = 2\langle i\delta_i \rangle - (N+1)\langle \delta_i \rangle \\ = \frac{2}{N-2}\sum_{i=h+1}^{N-1} i\delta_i + \frac{N+1}{N-2}\sum_{i=h+1}^{N-1}\delta_i \quad (9)$$

which describes the overall deviation from the mean atomic radius, as discussed in the following. Note that the summations in Eq. (9) run over all species except the smallest and largest species, indexed as $h$ and $N$ in Fig. 1.

Comparing Eqs. (7) and (8) indicates that a small $\delta$ value, as for some NOT-SS alloys in Figs. 2(a) and (b), can result from either having a small $\Delta r$ and/or having a small deviatoric energy term, i.e., the second term in Eq. (9). The latter is more likely for NOT-SS HEAs as the $\Delta r$ (being proportional to $\lambda^*$) for these alloys is found to be usually significant (Fig. 2). Figure 4 shows the ratio of the two terms in Eq. (8) for AM and IM forming alloys studied in Fig. 2. The results show that indeed for the IM and AM HEAs a small $\delta < 0.06$ corresponds to a small $\frac{\xi}{\Delta r}$ value, as marked in Fig. 4.

The $\xi$ parameter provides information about the *atomic-size distribution* of a given alloy and its effect on the elastic energy. In Eq. (9), any deviation $\delta_i$ will be weighted depending on its $i$ index. For $i < \frac{N+1}{2}$, the $\delta_i$ will be weighted negatively, for $i > \frac{N+1}{2}$ it will be weighted positively and for $i = \frac{N+1}{2}$, if exist, $\delta_i = 0$. For example, for 7- and 8-componet equiatomic alloys we get

$$\xi_7 = \frac{1}{5}(-4\delta_2 - 2\delta_3 + 2\delta_5 + 4\delta_6) \\ \xi_8 = \frac{1}{6}(-5\delta_2 - 3\delta_3 - \delta_4 + \delta_5 + 3\delta_6 + 5\delta_7). \quad (10)$$

From Eqs. (9) and (10) it becomes clear that a greater $\xi$ value indicates a larger atomic-size deviation from the mean atomic radius. This is illustrated in Fig. 5, demonstrating two extreme scenarios. For smaller $\xi$ values, the atomic-sizes are rather converging to the mean atomic radius while large $\xi$ values give a broader atomic-size distribution. When $\xi$ is large enough a bimodal atomic-size distribution forms, with its peaks close to the smallest and/or largest species, Fig. 5. Using Kube and Schroers's terminology [27], a low-$\xi$ alloy can be also called a 'homo-disperse' alloy while a high-

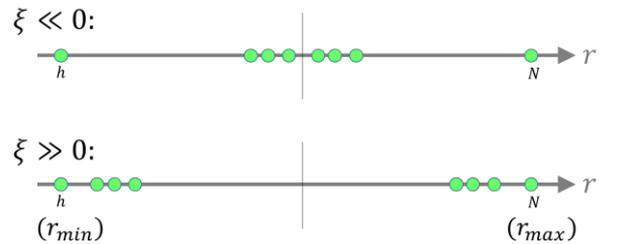

**Figure 5**: The atomic-size distribution as a function of $\xi$ parameter. Two extreme atomic-size distribution scenarios are shown. A small $\xi$ value results in a narrow size distribution close to the mean atomic-size while a large $\xi$ value results in a spread, bimodal atomic-size distribution. The $\xi$ parameter affects the elastic energy (the $\delta$ parameter) according to Eq. (9).

$\xi$ alloy is a 'hetero-disperse' alloy, in terms of their atomic-size distributions.

To understand the effect of atomic-size distribution ($\xi$ parameter), two IM forming HEAs are exemplified in Fig. 6(b) and (c): While the $\lambda^*$ values are similar (~0.17), the equiatomic TiVCrCuFeMnCoNi [28] gives relatively small values $\frac{\xi}{\Delta r} = -0.901$ and $\delta = 0.055$ (below the 0.06 threshold), whereas the Al$_{0.5}$CoCrCuFeNiTi$_x$ alloys (with x = 0.8-2) [29] give $\frac{\xi}{\Delta r} = +0.059$ and $\delta = 0.063$-$0.073$. We find that the absence of the Mn and V in in the latter case results in heavily uneven atomic-size distribution (hetero-disperse size distribution) and therefore larger $\frac{\xi}{\Delta r}$ and $\delta$ values for the Al$_{0.5}$CoCrCuFeNiTi$_x$ alloys compared to the TiVCrCuFeMnCoNi alloy, Figs. 6(b) and (c).

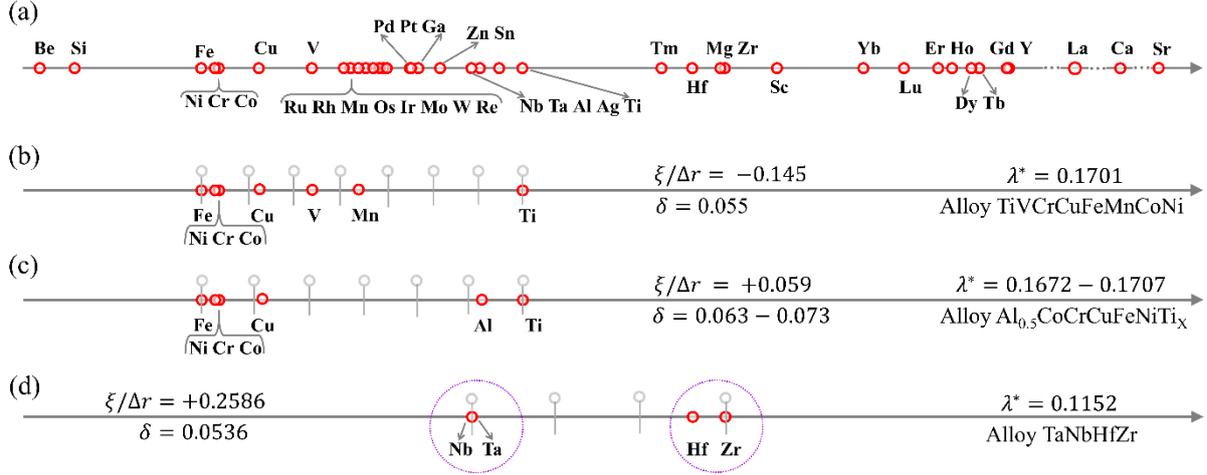

**Figure 6**: Atomic-size distributions of alloying elements. (a) The relative atomic-sizes of all species forming the alloys considered in this study. (b) TiVCrCuFeMnCoNi [28], (c) Al$_{0.5}$CoCrCuFeNiTi$_x$ [29] and (d) TaNbHfZr [30] alloys. (b) and (c) are IM forming with $\lambda^* > 0.16$ but different $\delta$ values below and above 0.06 and (d) is a solid solution forming alloy with $\lambda^* < 0.16$ and $\delta > 0.06$. The Nb-Ta and Hf-Zr atoms gives a large $\xi$ value and tend to form short-range ordering in TaNbHfZr [30].

Screening the available species for composing HEAs, Fig. 6(a), many combinations of species can be thought forming homo- or hetero-disperse size distributions. Although the possible combinations of atomic-sizes are limited by other physical effects, it is still conceivable that the elastic stability of an alloy mixture can be tailored by the choice of atomic-size distribution and parameters $\lambda^*$, $\delta$ (or $e$) and $\xi$, Eqs. (4)−(6), (8) and (9).

Here the $\lambda^*$ index turns out to be the primary parameter reflecting on the solubility and phase stability of a multi-component solid solution. A lower $\lambda^*$ value indicates a lower probability for a chemical decomposition. On the other hand, for a given solid solution we showed that the $\delta$ parameter is directly linked to the mixing elastic energy $e$. This information can be now immediately applied in the theories of multi-component solid-solution strengthening, where the $\delta$ parameter is used as an input parameter [31,32].

We have shown that the $\lambda^*$ and $\delta$ parameters can be rather independent. A combination of small $\lambda^*$ and $\delta$ values is likely to result in a stable solid solution with a low elastic energy while a combination of large $\lambda^*$ and small $\delta$ values was explained by having a homo-disperse size distribution (small $\xi$ value, Fig. 5). A combination of a small $\lambda^*$ and a large $\xi$ value can be also very interesting: When $\xi$ is large, i.e., having a hetero-disperse atomic-size distribution, it can drive formation of short-range ordering. This is, for example, evidenced in TaNbHfZr [30], illustrated in Fig. 6(d). For this alloy, $\frac{\xi}{\Delta r} = +0.259$, while $\lambda^* = 0.115$ and $\delta = 0.0536$. Here the Nb-Ta and Hf-Zr pairs form a bimodal hetero-disperse atomic-size distribution (Fig. 6(d)) with a large $\xi$ value (Fig. 5). The hetero-disperse atomic-size distribution then results in elastically-driven exchange between species leading to the formation of short-range ordering in the alloy. Here replacing Nb/Ta around Hf/Zr (and vice versa) have the largest elastic driving force, Eq. (5), with $\lambda_{min} \approx 0$, and lead to the clustering of Nb-Ta and Hf-Zr pairs, that have similar atomic-size.

As a final remark, we discuss now the effect of the number of constituting elements $N$ on the elastic energy. Originally, it has been suggested that maximizing the configurational entropy can be a key approach for stabilizing solid-solution HEAs. With increasing $N$, the configurational entropy of an equiatomic HEA increases as $\Delta S^{\text{conf}} = -R \sum_1^N \frac{1}{N} \ln\left(\frac{1}{N}\right)$, and the Gibbs free energy decreases with $-T\Delta S^{\text{conf}}$, where $T$ is the temperature. Similar to the configurational entropy, we found that the elastic energy of an $N$-component equiatomic alloy also rapidly decreases with increasing $N$, Eq. (8), provided that the quantities $q$, $r_{min}$ and $\Delta r$ remain similar. Here the two elastic and entropic effects add, both decreasing the Gibbs free energy of the alloy. One finds that the rate of the decrease in the elastic energy is greater than the rate of decrease in the configurational entropy: For the first term in Eq. (8)),

$$\frac{de}{dN} \propto \frac{1}{(N-1)^2} \ll \frac{d(-T\Delta S^{\text{conf}})}{dN} \propto \frac{1}{N}. \quad (11)$$

This means that depending on the coefficients $\frac{q \Delta r^2}{12\left(r_{min}+\frac{\Delta r}{2}\right)^2}$ vs. $-\frac{RT}{V_m}$, the leading dependency of the Gibbs free energy of mixing on the $N$ may switch between entropic and elastic effects. For instance, for $N = 4$, $q = 200$ GPa, $\Delta r = 0.03$ nm, $r_{min} = 0.13$ nm and a molar volume $V_m = 10^{-5}$ m$^3$ per mole (data for NbTiVZr [33]) we have $\frac{de}{dN} = \frac{d(-T\Delta S^{\text{conf}})}{dN}$ at $T \approx 400$ K. Hence, for any lower temperature the $N$-dependence of elastic energy dominates over the entropy effect.

In summary, we derived here the elastic energy of mixing for multi-component alloys and study phase stability in HEAs. Based on our elastic energy formulation, we investigated the tendency of an alloy for chemical decomposition and found a strain index $\lambda^* = \sqrt{\lambda_{max}^2 - \lambda_{min}^2}$. We show that a threshold of $\lambda^* < 0.16$, close to the Hume-Rothery limit for binary solid solutions, has a remarkable ability in identifying SS multi-component alloys, far better than the well-known parameter $\delta$.

We also derived $e = q\delta^2$, establishing a direct link between the elastic energy of mixing and the $\delta$ parameter. For equiatomic HEAs, the effect of the atomic-size distribution of constituting elements is captured by a new parameter, $\xi$. We found that a NOT-SS HEA ($\lambda^* > 0.16$) can have a low elastic energy (a small $\delta$ value) when the atomic-size distribution is homo-disperse, converging towards the mean atomic-size, quantified by a small $\xi$ value. On the other hand, a large $\xi$ indicates a hetero-disperse, potentially bimodal atomic-size distribution, Fig. 5, which can promote formation of short-range ordering from a SS forming HEA.

Detailed analysis of specific HEAs can be the subject of future studies based on our current derivations. Obviously, other Hume-Rothery rules beyond the atomic-size rule are important as well and must be considered due to their additional effects on the solubility of alloys [34]. Nevertheless, the current derivations and discussions provide a new perspective on the elasticity in multi-component solid solutions. Based on the current results, a more complete assessment of the excess Gibbs free energy (composed of the mixing elastic energy and chemical enthalpy of mixing) is in reach, expanding the development of thermodynamic databases [19,35,36] and computational exploration [37–40] of the HEAs. Furthermore, the current description of the elastic energy $e$ can be immediately used in the theories of strengthening in HEAs [31,32,41,42]. Finally, one should note that the current derivations are based on Vegard's law in the linear elasticity limit in a solid solution, as discussed by Eshelby. Hence, any deviation from Vegard's law or linear elasticity should be additionally considered.

## Acknowledgments

We thank professor C.T. Liu and professor Z. Wang for sharing their superalloys' composition data. The current research is being conducted within the project DA 1655/2-1 in the Heisenberg programme of the German Research Foundation (DFG). RDK gratefully acknowledges the financial supports within this programme and in project DA 1655/1-2 within SPP1713.

## Data availability

The data from this study are available per reasonable request.

## Competing financial interests

The authors declare no competing financial interests.

## Appendix A: Elastic energy of mixing for multi-component solid solutions

In his seminal work [18], Eshelby showed that the elastic energy of a binary solid solution ($i$ solute in $h$ solvent) is

$$e = \frac{\gamma_h(\gamma_h - 1)}{\gamma_i(\gamma_i - 1)} q_{ih}\lambda_{ih}^2 X_B - q_{ih}\lambda_{ih}^2 X_i^2 \quad (A1)$$

with $q_{ih} = \frac{9K_i(\gamma_h-1)}{2\gamma_h}$, $\gamma_{i/h} = \frac{3K_{i/h}+4\mu_h}{3K_{i/h}}$ and, $K_i$ and $\mu_i$ the bulk and shear moduli of substance $i$. Equation (A1) is based on the so-called sphere-in-hole model, considering a random binary solid solution in the limit of linear elasticity on the continuum level. In this derivations Eshelby considers substitution of a spherical volume made of one $i$ atom into a hole generated by removing one $h$ atom. The first term in this relation is the (self-)energy due to the substitution and the second term arises from the elastic interaction among the solute atoms. Here $q_{ih}\lambda_{ih}^2 = \frac{K_i\Delta V_i\Delta V_i^I}{2V_h^2}$ in which $V_h$ is the atomic volume of $h$ solvent and we have the total volume change $\Delta V_i = \Delta V_i^\infty + \Delta V_i^I$ where $\Delta V_i^\infty$ is the volume change when the matrix is infinite and $\Delta V_i^I$ is the correction due to free surfaces (here the surface of other inserted solute atoms). Note that the volume changes are obtained by solving the basic mechanical equilibrium. The following equalities are valid:

$$\Delta V_i = 3V_h\lambda_{ih} \quad (A2)$$

$$\Delta V_i^\infty = \Delta V_h = \frac{3V_h\lambda_{ih}}{\gamma_h} \quad (A3)$$

$$\Delta V_i^I = 3V_h\lambda_{ih}\frac{\gamma_h - 1}{\gamma_h} \quad (A4)$$

In order to determine the elastic energy of a multi-component solid solution, the sphere-in-hole model must be generalized to account for the interaction among dissimilar $i$ and $j$ solute atoms. The procedure to obtain the interaction energy between defects is to determine the interaction of one defect with the elastic field generated by another. On the continuum level, this will be equal to the work done by the pressure $p$ due to the 'already existing' solute atoms on the 'newly added' ones.

Along with the formation of the solution mixture, the interaction energy grows progressively assuming that the distribution of solutes is random/homogeneous. For one solute atom added, the pressure is equal to

$$p_1 = -\frac{K_i\Delta V_i^I}{N_0 V_h} = -\frac{2}{3}\frac{q_{ih}\lambda_{ih}}{N_0}. \quad (A5)$$

If we sum up all additional $i - i$ interaction energies for adding $n_i$ solute atoms, we obtain

$$e_{ii}^{int} = \frac{1}{N_0 V_h}\sum_{k=1}^{n_i} p_k \Delta V_i = -q_{ih}\lambda_{ih}^2 X_i^2 \quad (A6)$$

which is the second term in Eq. (A1) with $X_i = \frac{n_i}{N_0}$. Here the summation is over all $n_i$ solute atoms and we applied $\sum_{k=1}^{n_j} k = \frac{n_j(n_j-1)}{2} \approx \frac{n_j^2}{2}$. Now, by adding one $j$ solute atom the interaction between the existing $i$ atoms and the new $j$ atom increase the energy with

$$p_{n_i}\Delta V_j = -2q_{ih}V_h\lambda_{ih}\lambda_{jh}\frac{n_i}{N_0} \quad (A7)$$

where we use $\Delta V_j = 3V_h\lambda_{jh}$ and the pressure in the system increases as

$$-\frac{2}{3}q_{ih}\lambda_{ih}\frac{n_i}{N_0} - \frac{2}{3}q_{jh}\lambda_{jh}\frac{1}{N_0}. \quad (A8)$$

Here the first term builds the cross-interaction energy while the second term accounts for the pressure introduced by the new $j$ atom, building into the $j - j$ interaction energies similar to Eq. (A6). For $n_j$ additional atoms, the cross-interaction energies can be written as

$$\begin{aligned} e_{ij}^{int} &= \frac{1}{N_0 V_h}\sum_{k=1}^{n_j} p_{n_i}.\Delta V_j \\ &= -2q_{ih}\lambda_{ih}\lambda_{jh}\frac{n_i n_j}{N_0^2} \\ &= -2q_{ih}\lambda_{ih}\lambda_{jh}X_i X_j \end{aligned} \quad (A9)$$

Obviously, the cross-interaction energy must be independent of the order of addition, i.e., if we reverse the procedure and add $i$ solute atoms to a solid solution containing $j$ solute atoms the ultimate energy must be the same. Thus, applying a geometrical averaging of the two possible scenarios, we can write

$$e_{ij}^{int} = -2\sqrt{q_{ih}q_{jh}}\lambda_{ih}\lambda_{jh}X_i X_j. \quad (A10)$$

By generalizing this procedure, we obtain the elastic energy of substitutional $N$-component solid solutions as

$$\begin{aligned} e &= \sum_{i \neq h}^{N} q_{ih}\lambda_{ih}^2 \frac{\gamma_h(\gamma_h-1)}{\gamma_i(\gamma_i-1)} X_i \\ &- \sum_{i \neq h}^{N} q_{ih}\lambda_{ih}^2 X_i^2 \\ &- 2\sum_{\substack{i,j \neq h \\ i \neq j}}^{N} \sqrt{q_{ih}q_{jh}}\,\lambda_{ih}\lambda_{jh}\,X_i X_j \end{aligned} \quad (A11)$$

in which the first sum is due to the self-energy of added solute atoms, the second sum is the energy due to the interaction between similar $(i - i)$ atoms and the third sum is due to the interaction among dissimilar $(i - j)$ atoms. Note that the first two sums run over all solute atoms and the third sum runs over all unrepeated pairs of $i$ and $j$s.

While Eq. (A11) includes anisotropy effects, in the current study we limit our analysis for elastically isotropic solid solutions. Assuming isotropic elastic constants with $\gamma_i s \equiv \gamma_h$ and $K_i s \equiv K_h$, we can simplify Eq. (A11) to

$$e = q \sum_{i \neq h}^{N} \lambda_{ih}^2 (X_i - X_i^2) \\ -2q \sum_{\substack{i,j \neq h \\ i \neq j}}^{N} \lambda_{ih} \lambda_{jh} X_i X_j \quad (A12)$$

with $q = \frac{9K_h(\gamma_h - 1)}{2\gamma_h} = 2\mu_h \frac{1+\nu_h}{1-\nu_h}$.

## Appendix B: Elastic driving forces for a chemical decomposition

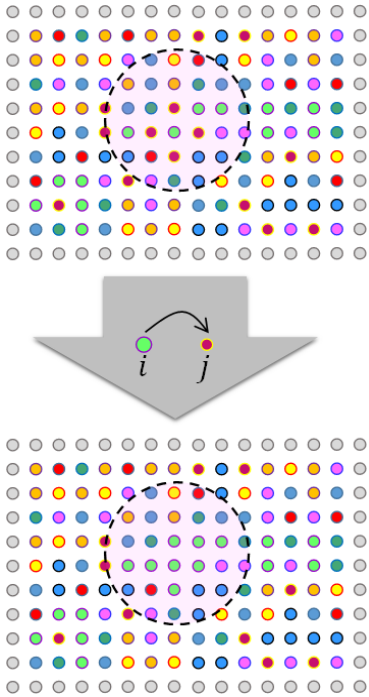

**Figure B1**: A binary $i - j$ substitution scenario within a small decomposition volume (marked by dashed circle), within an (infinite) matrix.

The total change in the elastic energy due to some compositional variations is given by

$$\Delta e = \sum_{i \neq h}^{N} \left(\frac{\partial e}{\partial X_i}\right)_{X_j} \Delta X_i \quad (B1)$$

and the constraint $\sum_i^N \Delta X_i = 0$ applies. In practice, an arbitrary number of elements can involve in a chemical decomposition process. The simplest scenario is a binary substitution of $j$ atoms with $i$ atoms. Considering such a binary decomposition scenario within a small decomposition volume (small compared to its parent homogeneous matrix solid solution), the composition/energy change in the matrix can be neglected. Figure B1 depicts this set-up. Thus, over the decomposition volume, with $\Delta X_j = -\Delta X_i$, we can write

$$\Delta e = e(X_i + \Delta X_i, X_j - \Delta X_i) - e(X_i, X_j) \\ = q\lambda_{ih}^2 (X_i + \Delta X_i - (X_i + \Delta X_i)^2) \\ + q\lambda_{jh}^2 \left(X_j - \Delta X_i - (X_j - \Delta X_i)^2\right) \\ -2q\lambda_{ih}\lambda_{jh} (X_i + \Delta X_i)(X_j - \Delta X_i) \quad (B2) \\ -2q\lambda_{ih}(X_i + \Delta X_i) \sum_{k \neq h,i}^{N} \lambda_{kh} X_k \\ -2q\lambda_{jh}(X_j - \Delta X_i) \sum_{k \neq h,j}^{N} \lambda_{kh} X_k$$

that gives

$$\left.\frac{\Delta e}{\Delta X_i}\right|_j = q(\lambda_{ih}^2 - \lambda_{jh}^2) \\ -2q(\lambda_{ih}^2 - \lambda_{ih}\lambda_{jh})X_i \\ +2q(\lambda_{jh}^2 - \lambda_{ih}\lambda_{jh})X_j \quad (B3) \\ -2q(\lambda_{ih} - \lambda_{jh}) \sum_{k \neq h,i,j}^{N} \lambda_{kh} X_k \\ + O(\Delta X_i^2).$$

If $\lambda_{ih} = \lambda_{jh}$ the elastic energy change vanishes. In other cases, depending on the magnitude of $\lambda_{ih}$ and $\lambda_{jh}$, the energy change can be either in favor (positive) or against (negative) the binary decomposition, i.e., enriching $i$ atoms against $j$ atoms. But since we have $\left.\frac{\Delta e}{\Delta X_i}\right|_j = -\left.\frac{\Delta e}{\Delta X_j}\right|_i$, one of the two enriching scenarios will be in favor of the chemical decomposition.

If assuming $X_i = X_j$, then we can further simplify

$$\left.\frac{\Delta e}{\Delta X_i}\right|_j = q(\lambda_{ih}^2 - \lambda_{jh}^2)(1 - 2X_i) \\ -2q(\lambda_{ih} - \lambda_{jh}) \sum_{k \neq h,i,j}^{N} \lambda_{kh} X_k \quad (B4) \\ + O(\Delta X_i^2).$$

If replacing $\lambda_{ih} = \lambda_{max}$ and $\lambda_{jh} = \lambda_{min}$ ($\lambda_{max} > \lambda_{min} > 0$) and considering that $\lambda^* = \sqrt{\lambda_{max}^2 - \lambda_{min}^2} \propto (\lambda_{max} - \lambda_{min})$, it becomes clear that $\left.\frac{\Delta e}{\Delta X_i}\right|_j \propto \lambda^*$ even if considering all terms in Eq. (B3).

It is also worth noting that a barrier to the phase decomposition arises due to the Cahn heterogeneous (coherent) elastic energy [31] (see [44] for more discussions). In the current binary decomposition scenario, the coherent elastic energy will be

$$q \sum_{i \neq h}^{N} \lambda_{ih}^2 \Delta X_i^2 - 2q \sum_{\substack{i,j \neq h \\ i \neq j}}^{N} \lambda_{ih} \lambda_{jh} \Delta X_i^2 \quad (B5)$$

Being second-order in $\Delta X_i$, however, this energy contribution becomes negligible compared to the energy change $\left.\frac{\Delta e}{\Delta X_i}\right|_j$, unless $\Delta X_i$ is significantly large.

Note that if the input parameters available, the exact value of $\left.\frac{\Delta e}{\Delta X_i}\right|_j$ is directly assessable for all pairs of species in any given alloy. The above procedure can be constructed for ternary or any higher-order chemical decomposition scenarios as well, resulting in more sophisticated criteria for chemical decomposition.

## Appendix C: Relationship between the mixing elastic energy and the $\delta$ parameter

For an $N$-component alloy, the $\delta$ parameter averages over the size differences of all $N$ species with respect to the mean atomic radius of the solution. On the other hand, the derivations in Appendix A show that the elastic energy of an $N$-component alloy is a function of the lattice distortion coefficients $\lambda_{ih}$s of $N-1$ elements with reference to the solvent element $h$. If rewriting the $\delta$ parameter in terms of $\lambda_{ih}$s, we obtain

$$\delta^2 = \sum_i^N X_i \left(\frac{r_i - \bar{r}}{\bar{r}}\right)^2$$

$$= \sum_i^N X_i \left(\lambda_{ih} + \frac{r_h - \bar{r}}{\bar{r}}\right)^2$$

$$= \frac{(r_h - \bar{r})^2}{\bar{r}^2}$$

$$+ \sum_{i \neq h}^N X_i \left(\lambda_{ih}^2 + 2\lambda_{ih}\left(\frac{r_h - \bar{r}}{\bar{r}}\right)\right) \quad (C1)$$

$$= \sum_{i \neq h}^N \lambda_{ih}^2 (X_i - X_i^2)$$

$$- 2 \sum_{\substack{i,j \neq h \\ i \neq j}}^N \lambda_{ih} \lambda_{jh} X_i X_j$$

$$= \frac{e}{q}$$

where we applied $\frac{r_h - \bar{r}}{\bar{r}} = -\sum_{i \neq h}^N X_i \lambda_{ih}$. Eq. C1 means that the conventional $\delta$ parameter is in fact a direct representation of the elastic energy under the assumption of elastic isotropy.

## Appendix D: Elastic energy of equiatomic multi-component solid solutions and the effect of atomic-size distribution

Equiatomic HEAs have been the basis for expanding the concept of multi-principal element alloys. This is particularly because the effect of '$N$' (number of constituting species) becomes intuitive having $X_i$s $= \frac{1}{N}$. The isotropic elastic energy of an $N$-component equiatomic alloy simplifies to

$$e = \frac{q(N-1)}{N^2} \sum_{i \neq h}^N \lambda_{ih}^2 \\ - \frac{2q}{N^2} \sum_{\substack{i,j \neq h \\ i \neq j}}^N \lambda_{ih} \lambda_{jh} . \quad (D1)$$

In principle, the strain coefficients $\lambda_{ih}$s also depend on the $N$, through the mean atomic radius $\bar{r}$. To include this dependency, we can explore the possible scenarios for constructing $\bar{r}$ and $\lambda_{ih}$s. To do so, we start with considering an ideal atomic-size distribution such that the atomic-sizes are equally distanced by $d = \frac{\Delta r}{N-1}$ with $\Delta r = r_{max} - r_{min}$. Figure 1(b) depicts this ideal configuration. Thus, for an equiatomic alloy $\bar{r} = r_{min} + \frac{\Delta r}{2}$ and $\lambda_{(h+k)h} = \frac{r_{h+k} - r_h}{\bar{r}} = \frac{r_{h+k} - r_{min}}{\bar{r}} = \frac{kd}{\bar{r}} = \frac{k \Delta r}{(N-1)\bar{r}}$ with $h$ chosen to be the smallest species in the alloy. For this ideal equidistance configuration, we obtain

$$e = \frac{q(N-1)d^2}{\bar{r}^2 N^2} \sum_{i \neq h}^N i^2 \\ - \frac{2q\, d^2}{\bar{r}^2 N^2} \sum_{\substack{i,j \neq h \\ i \neq j}}^N i \cdot j . \quad (D2)$$

Computing the numerical sums and expanding the parameters $d$ and $\bar{r}$,

$$e = \frac{q\, \Delta r^2}{12\left(r_{min} + \frac{\Delta r}{2}\right)^2} \left(\frac{N+1}{N-1}\right). \quad (D3)$$

To achieve the elastic energy of a real equiatomic alloy with arbitrary atomic-sizes, a correction must be introduced to Eq. (D3) due to the deviations from the ideal equidistance configuration. To further our derivation, we define deviatoric atomic mismatches $\delta_{h+k} = r_k - kd$, as illustrated in Fig. 1(c). Whether the atomic radius $r_i$ is larger or smaller than its ideal equidistance radius $= id$, $\delta_i$ can be positive or negative. We have then $\lambda_{(h+k)h} = \frac{(h+k-1)d + \delta_{h+k}}{\bar{r}}$ and the additional deviatoric energy terms can be derived as:

$$\frac{2q(N-1)d}{\bar{r}^2 N^2} \sum_{i=h+1}^{N-1} (i-1)\delta_i \\ - \frac{2qd}{\bar{r}^2 N^2} \sum_{i=h+1}^{N-1} \left(\frac{N(N-1)}{2} - (i-1)\right)\delta_i \quad (D4) \\ + O(\delta_i^2)$$

$$= \frac{2qd}{\bar{r}^2 N} \sum_{i=h+1}^{N-1} i\delta_i$$

$$- \frac{2qd}{\bar{r}^2 N}\left(\frac{N+1}{2}\right) \sum_{i=h+1}^{N-1} \delta_i + O(\delta_i^2)$$

$$= \frac{q\,\Delta r\,[2\langle i\delta_i\rangle - (N+1)\langle\delta_i\rangle]}{N\left(r_{min}+\frac{\Delta r}{2}\right)^2}\left(\frac{N-2}{N-1}\right)$$

$$+ O(\delta_i^2).$$

Here we have introduced two averaged quantities $\langle i\delta_i\rangle = \frac{1}{(N-2)}\sum_{i=h+1}^{N-2} i\delta_i$ and $\langle\delta_i\rangle = \frac{1}{(N-2)}\sum_{i=h+1}^{N-2}\delta_i$. Inserting Eq. (D4) into Eq. (D3) gives

$$e = \frac{q\,\Delta r^2}{12\left(r_{min}+\frac{\Delta r}{2}\right)^2}\left(\frac{N+1}{N-1}\right)$$

$$+ \frac{q\,\Delta r\,\xi}{N\left(r_{min}+\frac{\Delta r}{2}\right)^2}\left(\frac{N-2}{N-1}\right) + O(\delta_i^2). \quad \text{(D5)}$$

The second term in Eq. (D5) can be positive or negative depending on the sign of the expression

$$\xi = 2\langle i\delta_i\rangle - (N+1)\langle\delta_i\rangle. \quad \text{(D6)}$$

The $\xi$ parameter implies that any deviation $\delta_i$ will be weighted depending on its $i$ index. The lower values of the $\xi$ parameter result in lowering the elastic energy in the equiatomic solid solution. Note that although the higher-order contributions, proportional to $\delta_i^2$s, are neglected in Eq. (D6), they might become significant depending on the magnitude of the $\delta_i$ values.

## Graphical Abstract

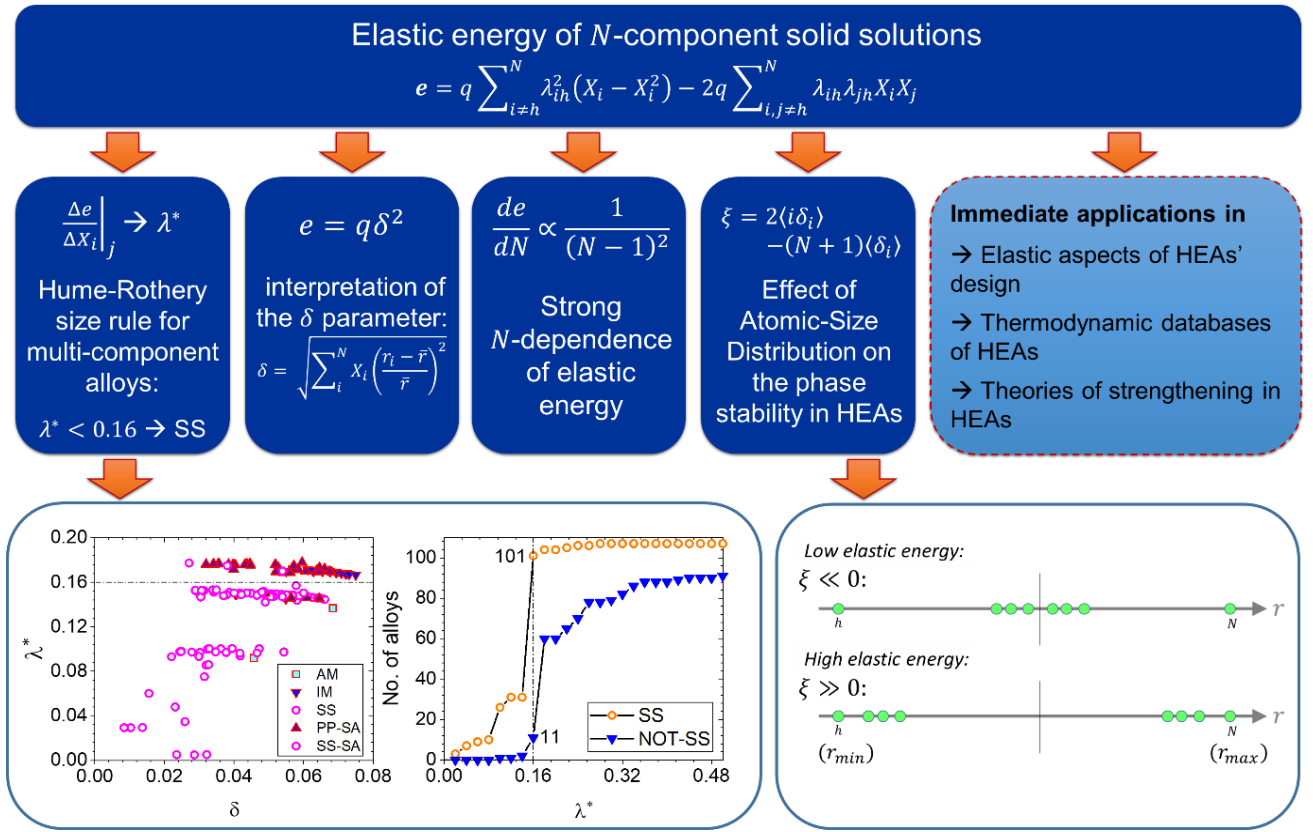